\documentclass[a4paper,fleqn,usenatbib]{mnras}

\usepackage{paralist,color}

\usepackage[T1]{fontenc}
\usepackage{ae,aecompl}
\usepackage[normalem]{ulem} 

\usepackage{graphicx}	
\usepackage{amsmath}	
\usepackage{amssymb}	

\title[The quiescent state of GRS\,1747--312]{The quiescent state of the neutron-star X-ray transient GRS\,1747--312 in the globular cluster Terzan\,6}

\author[S. Vats et al.]{Smriti Vats$^1$,\thanks{E-mail: s.vats@uva.nl}
Rudy Wijnands$^1$,
Aastha S. Parikh$^1$,
Laura Ootes$^1$,
Nathalie Degenaar$^1$,
\newauthor Dany Page$^2$
\\
$^1$ Anton Pannekoek Institute for Astronomy, University of Amsterdam, Science Park 904, 1098 XH Amsterdam, The Netherlands\\
$^2$ Instituto de Astronom\'{i}a, Universidad Nacional Aut\'{o}noma de Me\'{x}ico, Mexico D.F. 04510, Mexico}

\date{Accepted XXX. Received YYY; in original form ZZZ}

\pubyear{2018}

\begin{document}
\label{firstpage}
\pagerange{\pageref{firstpage}--\pageref{lastpage}}
\maketitle

\begin{abstract}

We studied the transient neutron-star low-mass X-ray binary GRS\,1747--312, located in the globular cluster Terzan 6, in its quiescent state after its outburst in August 2004, using an archival {\it XMM-Newton} observation. A source was detected in this cluster and its X-ray spectrum can be fitted with the combination of a soft, neutron-star atmosphere model and a hard, power-law model. Both contributed roughly equally to the observed 0.5-10 keV luminosity ($\sim4.8\times10^{33}$ erg\,s$^{-1}$). This type of X-ray spectrum is typically observed for quiescent neutron-star X-ray transients that are perhaps accreting in quiescence at very low rates. Therefore, if this X-ray source is the quiescent counterpart of GRS\,1747--312, then this source is also accreting at low levels in-between outbursts. Since source confusion a likely problem in globular clusters, it is quite possible that part, if not all, of the emission we observed is not related to GRS\,1747--312, and is instead associated with another source or conglomeration of sources in the cluster. Currently, it is not possible to determine exactly which part of the emission truly originates from GRS1747-312, and a {\it Chandra} observation (when no source is in outburst in Terzan 6) is needed to be conclusive. Assuming that the detected emission is due to GRS\,1747--312, we discuss the observed results in the context of what is known about other quiescent systems. We also investigated the thermal evolution of the neutron-star in GRS\,1747--312, and inferred that GRS\,1747--312 can be considered a typical quiescent system under our assumptions.
\end{abstract}




\section{Introduction}\label{section:ch4-intro}

Low-mass X-ray binaries (LMXBs) are binary systems harbouring a primary compact accretor (i.e., black hole or neutron-star) and a secondary donor, which is typically a normal star with mass $M \lesssim 1 ~ M_{\odot}$. Mass transfer from the donor occurs via Roche-lobe overflow. The matter forms an accretion disk around the primary star before it is accreted onto the compact accretor. It is this accretion that leads to the observed X-ray emission in these systems \citep[ see, e.g., the review chapters in][for more information about X-ray binaries]{1997xrb..book.....L,2010csxs.book.....L}.

Several tens of LMXBs are persistent sources, which means that they are always accreting at relatively high accretion rates (typically producing X-ray luminosities of $L_{X}>10^{35-39}$ erg s$^{-1}$), making them easy to detect and study with currently available X-ray telescopes. However, most systems are what we call transient LMXBs (or X-ray transients) that are typically very faint (i.e., they are in quiescence) but occasionally exhibit bright X-ray outbursts. These outbursts are due to a sudden, large increase in the rate of accretion onto the compact accretor, causing a similarly large increase in their X-ray luminosities. Transient LMXBs go into outbursts for periods lasting typically from weeks up to months, although in some rare cases these outbursts can last years up to even several decades. During their outbursts the transients can reach X-ray luminosities of $L_X \sim10^{35-39}$ erg s$^{-1}$, which is very similar to what is observed from the persistent LMXBs (in outburst the transients indeed look very similar to the persistent systems in many aspects). During quiescence, these systems either accrete at a very low level or they do not accrete at all, and have X-ray luminosities of $L_X \sim 10^{30-34}$ erg s$^{-1}$.

When neutron-star X-ray transients \citep[for a review of such systems, see][]{campana98} are in their quiescent state, their X-ray spectra in the 0.5--10~keV energy range can typically have two components (although with a large variety in relative strengths): a soft black-body like component dominating the spectra below 1--2~keV, and a hard power-law shaped component at higher energies \citep[e.g., see][ and the articles referencing these papers]{1996PASJ...48..257A,1998PASJ...50..611A,2001ApJ...551..921R,2001ApJ...559.1054R,rutledge02a}. The soft spectral component can be attributed to thermal radiation that very likely originates from the surface of the neutron-star. This radiation could be due to the heat generated by accretion onto the surface at very low rates \citep[e.g.][]{zampieri95}, or due to internal heat from the neutron-star itself that is released at the surface \citep[ e.g.][]{brown98}. The power-law component of a quiescent neutron-star LMXB may originate from residual accretion of matter onto the neutron-star surface or magnetosphere, however, currently our understanding of this component is inadequate \citep[see the discussions in, e.g.,][]{campana98,degenaar+12,chakrabarty14}.

During an outburst, the accreted matter compresses the neutron-star crust and heats it up through a set of nuclear reactions (viz. electron captures, neutron emissions, and pycnonuclear reactions), resulting in the release of $\sim$1--2~MeV of energy per accreted nucleon \citep{sato79,haenselzdunik90,haenselzdunik03,haenselzdunik08,steiner12}. The heat generated by these reactions typically flows into the core, with a small fraction also transported upwards to the surface \citep{brown98}. Due to these processes, the crust and the core go out of thermal equilibrium with one another (if the accretion outburst last for a long period of time -- typically many months to years). When the X-ray transient returns back into quiescence, the hot crust cools down by conducting this excess heat to the core and to the surface until it attains thermal equilibrium with the core again \citep{rutledge02}. Hence, after many repeated accretion episodes, the core too gets significantly heated (e.g., see \citealt{colpiea01,wijnandsea13}). By studying the heating of the crust, and its subsequent cooling, we can investigate the physics at work in the very dense environments of neutron-star crusts \citep[for a recent observational review see][]{wijnands17}. Also, the longer we study a neutron-star LMXB in quiescence, the deeper we are able to probe into its layers. Systems for which such studies are possible typically show quiescent spectra that are dominated by the soft spectral component \citep[e.g.,][]{2013ApJ...774..131C,2013ApJ...775...48D,2014ApJ...791...47D,homanea14,2016ApJ...833..186M,2017ApJ...851L..28P}, although sometimes a significant power-law component is present as well \citep[e.g.,][]{fridriksson2010,fridriksson2011,degenaar15b,2017MNRAS.466.4074P,2018MNRAS.tmp..408P}.

For systems having a quiescent spectrum almost fully dominated by a hard component \citep[typically fitted with a power-law model with photon index $\Gamma$ of 1-2; e.g. ][]{2002ApJ...575L..15C,2005ApJ...618..883W,2007ApJ...660.1424H,2009ApJ...691.1035H,degenaar+12}, the quiescent emission might arise due to residual accretion onto the neutron-star magnetosphere. Alternatively, the emission might be due to shocks arising from interaction of the pulsar wind (assuming that the radio pulsar mechanism becomes active) with inflowing matter from the companion star. However, despite those tentative interpretations, the origin of the power-law component in these quiescent systems remains rather unclear (for discussions see \citealt{campana98,degenaar+12,degenaarwijnands12}). 

There also exist a group of systems where the soft and hard components contribute nearly equally to the quiescent spectrum (typically measured in the 0.5--10 keV energy range). In such cases as well, the origin of the power-law component is not fully understood. Moreover, it is unclear if this power-law component is due to the same mechanisms as the one seen for the power-law dominated systems or if it is due to different processes. However, recent evidence suggests that in sources where both soft and hard components are nearly equally strong, both may be occurring because of low-level accretion onto the neutron-star surface \citep{chakrabarty14,dangelo15,wijnands15}.

\subsection{GRS 1747--312} \label{subsection:ch4-grs1747}

\begin{figure*}
\includegraphics[clip=,width=2.0\columnwidth]{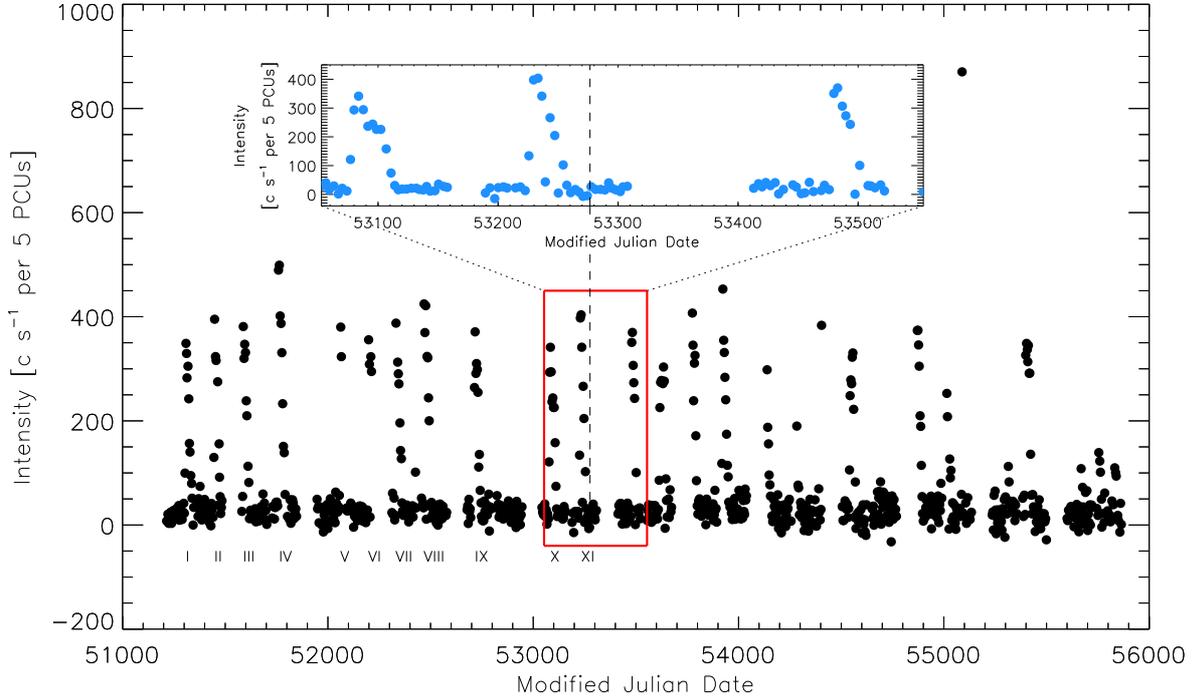}
\caption{ The {\em RXTE}/PCA (2--60~keV) light curve of GRS\,1747--312 obtained between February 1999 and October 2011 as part of the Galactic Bulge Scan program \citep{swankmarkwardt01}. The time when the {\it XMM-Newton} quiescence observation was performed is shown by the dashed line (i.e., it was performed after outburst XI). The roman numbering below the light curve indicates the outbursts that we used to model the thermal evolution of the neutron-star (Sect.\,\ref{subsection:ch4-cooling}) }
 \label{ch4_lc}
\end{figure*}

GRS\,1747--312 is a transient LMXB in the globular cluster Terzan\,6. It was first discovered in 1990 with the ART-P instrument on-board {\em Granat} \citep{pavlinsky+94} and with {\em ROSAT} \citep{predehl+91}. The source is located at $\alpha_\text{J2000}=17^{h}50^{m}46.862^{s}$ and $\delta_\text{J2000}=-31^{o}16\arcmin28\farcs86$, with a 95\% error radius of 0\farcs4. This position was obtained by \citet{intzand03} using an archival $Chandra$ observation (ObsID 720) of the source in outburst. The discovery of type-I X-ray bursts from GRS\,1747--312 demonstrated that the accretor in this system is a neutron-star \citep{intzand03}. This source is quite remarkable in the fact that its outbursts recur very often (see, eg., Fig.\,\ref{ch4_lc}) -- with an average recurrence time of 136 days -- and has similar outburst behaviour during all observed outbursts 
\citep{intzand03,simon09}. There are only a handful of other sources that exhibit similar behaviour, with the best known being the Rapid Burster in the globular cluster Liller~1 (e.g., \citealt{masetti02,simon08}), Aql~X-1 (e.g., \citealt{maitrabailyn08,campanaea13,gungorea14}), and H\,1743--322 (e.g., \citealt{altamiranostrohmayer12,zhouea13}; a black-hole transient that shows very regular outbursts since 2003). GRS\,1747--312 is an eclipsing source with an orbital period of $\sim$0.5 d \citep{intzand03}.

Terzan\,6 was observed using {\it XMM-Newton} in 2004 when GRS\,1747--312 was in quiescence, and the results were reported by \citet{saji16} in conjunction with results from observations during its outburst state (using $Suzaku$, $Chandra$ and $Swift$). However, they did not consider potential source confusion due to other unrelated sources in Terzan 6 and did not use the standard spectral decomposition (for quiescent neutron-star LMXBs) of the {\it XMM-Newton} spectrum in their work (they used a "blackbody+CompTT" model). In addition, they did not put the quiescent properties of the source (if indeed the emission was due to GRS\,1747--312) into the context of quiescent neutron-star LMXBs, and neutron-star crust heating and cooling. Hence, we reexamined the {\it XMM-Newton} data and discuss the result of this observation in the above mentioned contexts.

\section{Observations and Data Analysis}

\begin{figure}
\label{ch4_pn-fov}
\includegraphics[clip=,width=1.0\columnwidth]{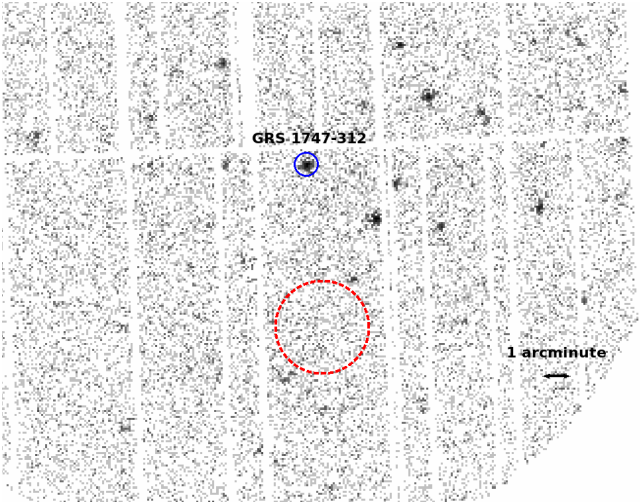}
\caption{The X-ray image of the field around GRS\,1747--312, as obtained using the EPIC-PN camera aboard {\it XMM-Newton}. The source region is marked with the blue circle with a radius of 25$\arcsec$ and the background region is marked with the dashed red circle with a radius of 100$\arcsec$.}
\end{figure}

GRS\,1747--312 was observed on September 28, 2004 using {\em XMM}-Newton (observation ID: 0206990101) with an exposure time of 8.2~ks for the PN and 14.3~ks each for the MOS detectors. The source was in quiescence during this observation (see Fig. \,\ref{ch4_lc}; see also \citealt{saji16}) and the data were obtained from the {\it XMM-Newton} Science Archive\footnote{\href{https://www.cosmos.esa.int/web/xmm-newton/xsa}{https://www.cosmos.esa.int/web/xmm-newton/xsa}} (XSA). We reduced the data using the Science Analysis System (SAS; version 15.0.0\footnote{\href{https://www.cosmos.esa.int/web/xmm-newton/sas}{https://www.cosmos.esa.int/web/xmm-newton/sas}}) using the standard analysis threads\footnote{\href{https://www.cosmos.esa.int/web/xmm-newton/sas-threads}{https://www.cosmos.esa.int/web/xmm-newton/sas-threads}}. Fig.\,\ref{ch4_pn-fov} shows the field of view of Terzan\,5 with GRS\,1747--312 indicated on it.

{\it XMM-Newton} collects data using the three EPIC cameras (PN and two MOS detectors), two RGS detectors, and the optical monitor (OM). However the source is too faint to be detected with the RGS, and lies in a crowded globular cluster field, making the optical OM data unavailing. Hence, in this work, we only used data obtained using the MOS1/2 and PN detectors. All EPIC cameras were used in full frame mode, using the medium filter. We processed the raw data using the tools {\tt emproc} and {\tt epproc} for the MOS1/2 and PN detectors, respectively. In order to obtain a data set free of background flaring, we examined the light curves in energies greater than 10~keV for MOS1/2 and between 10 and 12~keV for PN and used those light curves to determine the parts of the observations during which strong background flares occur. Those intervals were excluded from the data. For MOS1/2 we excluded intervals where the count rate was $> 0.15$ counts s$^{-1}$, and for PN where the count rate was $> 0.4$ counts s$^{-1}$. The resulting effective exposure times were 12, 11 and 7 ks for the MOS1, MOS2 and PN detectors, respectively. 

Since, GRS\,1747--312 is an eclipsing source, we investigated if any eclipse occurred during the {\em XMM} observations. We obtained the ephemeris for the eclipses from \citet{intzand03}, and found that the source did not exhibit an eclipse during the observation.

We extracted the spectra making use of the {\tt xmmselect} tool, using a circular source extraction region (centred on the source position as given by \citealt{intzand03}; see also Section\,\ref{subsection:ch4-grs1747}) having a radius of 25$\arcsec$. Circular background extraction regions were used with radii of 100$\arcsec$, placed over a source-free part of the CCD on which the source is located. We used the same source region for PN and MOS1/2, but different background regions for each of the detectors (see Figure\,\ref{ch4_pn-fov} for the source and background regions for the PN). We used {\tt rmfgen} and {\tt arfgen} to generate the redistribution matrices and ancillary response functions. We grouped the spectra using {\tt specgroup} such that each bin contained a minimum of 15 counts.

To study the long term evolution of the source around the time of the {\it XMM-Newton} observation, we used the {\em Rossi X-ray Timing Explorer} ({\em RXTE}) Proportional Counter Array (PCA) light curve of the source (see Fig. \ref{ch4_lc}) obtained as part of the Galactic Bulge Scan program\footnote{\href{https://asd.gsfc.nasa.gov/Craig.Markwardt/galscan/}{https://asd.gsfc.nasa.gov/Craig.Markwardt/galscan/}} \citep{swankmarkwardt01}.

\begin{figure}
\includegraphics[clip=,width=0.7\columnwidth,angle=270]{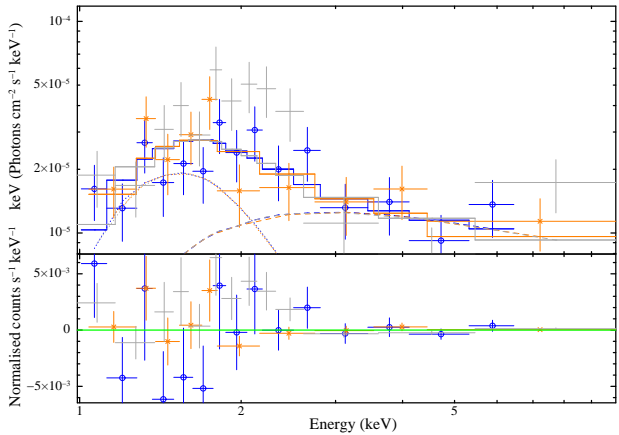}
\caption{The combined X-ray spectra obtained during the {\it XMM-Newton} observation -- PN($\circ$) in blue, MOS1($\times$) in orange and MOS2($+$) in grey. Solid lines represent the best fit model for the {\it XMM-Newton} data; the individual {\em nsatmos} and {\em powerlaw} models are represented by dotted and dashed lines respectively. As can be seen, the MOS2 spectrum (grey points) shows an excess of counts at around 2~keV, compared to the PN and MOS1 spectra. The bottom panel shows the fit residuals.}
\label{ch4-spectra}
\end{figure}

\section{Results}\label{section:ch4-results}

The grouped spectra of all three EPIC detectors (MOS1/2 and PN) were fit simultaneously using {\tt XSpec} (version 12.8.1), in the energy range 0.5--10~keV. Since the source was observed in quiescence (and assuming that the detected emission indeed comes from GRS\,1747--312; but see Section \ref{subsection:ch4-confusion}), we first tried to fit the spectra with the commonly used spectral model for quiescent neutron-star LMXBs, viz., a neutron-star atmosphere model ({\em nsatmos}; \citealt{heinke06}) affected by interstellar absorption modelled using {\em tbabs}; using {\tt WILM} abundances \citep{wilms00} and {\tt VERN} cross-sections \citep{verner96}. For the {\em nsatmos} model, we assumed a neutron-star of mass 1.4~M$_{\odot}$ and a radius of 10~km, at a distance of 9.5~kpc \citep{barbuy97,kuulkers03}. Also, we assumed that the whole surface of the neutron-star was emitting, so we set the fraction of the surface that is emitting (i.e., normalisation) to 1. This {\em nsatmos} model assumes that we observe incandescent thermal emission from the neutron-star surface. 

However, when using this model we obtained an unsatisfactory fit -- $\chi^{2}/\nu=3.49$ for 36 degrees of freedom (dof). To improve the fit, we added a power-law component ({\tt XSpec} {\em powerlaw} model; also commonly used in modelling quiescent spectra) to the model which resulted in a significant improvement of the fit ($\chi^{2}/\nu=1.38$ for 34 dof). We also performed an F-test to determine if the improvement in the fit was a chance occurrence. The probability of a chance occurrence in improvement of the fit is 1.3$\times10^{-7}$, which means that the addition of the power-law component indeed resulted in a model that could explain the spectral shape better. We then included the convolution model {\em cflux} to obtain the unabsorbed flux contributions (and their errors) for both components in the 0.5--10~keV energy range. We also infer the effective neutron-star temperature as seen by an observer at infinity, $kT_{\text{eff}}^{\infty}$ \footnote{$kT_{\text{eff}}^{\infty}=kT_{\text{eff}}/(1+z)$, where $(1+z)$ is the gravitational redshift factor, and is equal to 1.31 for our assumedneutron-star of mass 1.4~M$_\odot$ and radius 10~km.}. 

In Table \ref{tab:ch4-specfit} we list the obtained spectral parameters. As can be seen from this table the errors on the fit parameters were quite large when leaving all parameters free. This is due to the relatively low number of photons in our spectrum. In order to better constrain some of these parameters, we fitted the same combination of models ({\em tbabs}, {\em nsatmos} and {\em powerlaw}) keeping different combinations of two model parameters fixed, viz., the neutral hydrogen column density to the source, $N_\text{H}=1.4 \times 10^{22}$ cm$^{-2}$ \citep{intzand00} and the power law index $\Gamma=1.5$; a value often found for quiescent neutron-star X-ray transients. Details of the model parameters can be found in Table \ref{tab:ch4-specfit}. The obtained neutron-star temperature for the combined {\em nsatmos} and {\em powerlaw} model (with the $N_\text{H}$ free but with the photon index fixed to 1.5, which we roughly find to be the value for our chosen models, as well, if we fixed the $N_\text{H}$ to $1.4 \times 10^{22}$ cm$^{-2}$; see Table \ref{tab:ch4-specfit}) was $\sim$112 eV and the total 0.5--10~keV X-ray luminosity was $\sim$4.8$\times$10$^{33}$ erg\,s$^{-1}$ (calculated at a distance of 9.5\,kpc). Using this model, we found that the \textit{nsatmos} component contributes $\sim$55.5\% and the \textit{powerlaw} contributes $\sim$45.5\% to the total (unabsorbed) 0.5--10 keV luminosity.

We found that while the spectra for MOS1 and PN appeared to be consistent with each other, MOS2 showed an excess of counts at around 2~keV (see Figure\,\ref{ch4-spectra}). This effect was observed regardless of the choice of source and background extraction regions. We contacted the {\it XMM-Newton} help desk to discuss with them the possibility that the MOS2 data set might be flawed in some way. However, they did not find anything unusual or problematic with this dataset and therefore they concluded that all data could be used. Nonetheless, we repeated the same steps as before whilst fitting only the MOS1 and PN spectra to see what difference it makes to the observed temperature (see Table \ref{tab:ch4-specfit} for the results of these fits as well). For the single {\em nsatmos} model we obtain a $\chi^{2}/\nu=3.65$ for 23 dof, which again was not an acceptable fit. The combination of a {\em nsatmos} and {\em powerlaw} model results in an improvement of the fit ($\chi^{2}/\nu=0.83$ for 21 dof), which is corroborated by the F-test results (probability of chance occurrence$=1.7\times10^{-7}$). For this model, we fixed the photon index to 1.5, as the data quality does not allow to meaningfully constrain all spectral parameters independently (see Table \ref{tab:ch4-specfit}). As can be seen in Table \ref{tab:ch4-specfit}, exclusion of MOS2 data from the spectral fitting leads to a lower inferred neutron-star surface temperature ($kT_\text{eff}^{\infty}$). 

\begin{table*}
	\centering
	\caption{The results obtained from our X-ray spectral analysis}
	\label{tab:ch4-specfit}
	\begin{tabular}{lccccc} 
		\hline
		$N_\text{H}$         &              	Flux$_{nsatmos}^a$   &    	$kT_\text{eff}^{\infty}$             &       	$\Gamma$               &    	Flux$_{powerlaw}^a$  & $\chi^{2}/\nu$\\
	($10^{22}$ cm$^{-2}$)	         &              	(10$^{-13}$ erg\,cm$^{-2}$\,s$^{-1}$)    &  (eV)  	             &       	               &    	(10$^{-13}$ erg\,cm$^{-2}$\,s$^{-1}$) & \\\hline
PN+MOS1+MOS2  &	& 	& &	&\\ \hline
1.9$^{+0.5}_{-0.6}$ &  	3.1$\pm0.4$ & 	118.2$^{+9.9}_{-22.6}$&	1.01$^{+1.05}_{-1.03}$                           &	1.9$\pm0.3$ & 1.375 (34 dof)\\ 
1.4 (fixed) &  	1.6$\pm0.3$ & 	101.9$^{+7.9}_{-41.4}$&	1.6$^{+0.8}_{-0.7}$                            &	2.2$\pm0.3$&1.403 (35 dof)\\ 
1.8$\pm0.4$ &  	2.4$\pm0.4$ & 	112.0$^{+8.4}_{-10.7}$ &	1.5(fixed)                             &	2.0$\pm0.3$&1.352 (35 dof)\\ 
1.4 (fixed) &  	1.7$\pm0.3$ & 	103.8$^{+3.7}_{-4.2}$&	1.5(fixed)                             &	2.1$\pm0.3$&1.368 (36 dof)\\ \hline
PN+MOS1 &  	& 	& &	&\\ \hline
1.5$^{+0.6}_{-1.0}$ &  	1.5$\pm0.4$ & 	101.0$^{+17.6}_{b}$&	1.6$^{+0.7}_{-0.9}$                           & 2.3$\pm0.3$&0.829 (21 dof)\\ 
1.4 (fixed) &  	1.4$\pm0.3$ & 	99.3$^{+8.9}_{-31.0}$&	1.6$\pm0.7$                            &	2.3$\pm0.3$&0.792 (22 dof)\\ 
1.5$^{+0.5}_{-0.7}$ &  	1.6$\pm0.4$ & 	102.7$^{+11.7}_{-21.7}$&	1.5(fixed)                             &	2.2$\pm0.3$&0.792 (22 dof)\\ 
1.4 (fixed) &  	1.5$\pm0.3$ & 	100.8$^{+4.6}_{-5.7}$&	1.5(fixed)                             &	2.2$\pm0.3$&0.760 (23 dof)\\ \hline 
\multicolumn{3}{p {3cm}}{}                                             \\       
\multicolumn{6}{l}{All reported uncertainties are at the 90\% confidence level. }\\
\multicolumn{6}{l}{$^a$ The fluxes are unabsorbed fluxes and for the energy range 0.5-10 keV}\\
\multicolumn{6}{l}{$^b$ The lower error did not converge}\\
	\end{tabular}
\end{table*}

\section{Discussion}\label{section:ch4-discussions}

Here we presented an archival {\it XMM-Newton} observation performed on the globular cluster Terzan 6 at a time when the transient neutron-star LMXB, GRS\,1747--312, was in its quiescent state. This observation was previously reported by \citet{saji16}. However, they assumed that the source they detected was in fact the quiescent counterpart of this transient, and did not consider the possibility that unrelated cluster sources could have contributed part, or even most, of the observed emission. In addition, they modelled the obtained X-ray spectrum with a two component model (i.e., a black-body plus the {\em CompTT} model) that are typically not used for quiescent neutron-star X-ray transients. In order to better understand the quiescent properties of GRS\,1747--312 we have reanalysed these data.

We confirmed that the spectrum cannot be fitted with a single component model, so as a final spectral model we used a neutron-star atmosphere model ({\em nsatmos}), combined with a power-law component ({\em powerlaw}). We obtained a neutron-star temperature of $\sim$112 eV and total 0.5--10~keV X-ray luminosity of $\sim$4.8$\times$10$^{33}$ erg\,s$^{-1}$, with both spectral components contributing roughly equally to the total 0.5--10~keV luminosity. All these results are similar to what is frequently found for other quiescent neutron-star X-ray transients. Although this suggests that the emission possibly originated from a quiescent neutron-star transient, and potentially from GRS\,1747--312, source confusion might be a serious issue. We discuss this possibility in Section \ref{subsection:ch4-confusion}. Even though we cannot conclusively answer what fraction, if any, of the emission originated from the quiescent counterpart of GRS\,1747--312, it is interesting to investigate what could be inferred from this observation if in the future (e.g., by using a high angular resolution observation with {\it Chandra}) it could be proven that indeed GRS\,1747--312 dominates the observed flux. Therefore, in the final two sections of our discussion we explicitly assume that all the emission is due to GRS\,1747--312, and we discuss the observed properties in the context of the two main explanations for the quiescent spectra of such systems: residual low-level accretion onto the neutron-star (Section \ref{subsection:ch4-accretion}) and cooling emission from a neutron-star that has been reheated in outburst due to the accretion of matter (Section \ref{subsection:ch4-cooling}).

\subsection{Potential source confusion}\label{subsection:ch4-confusion}

In many Galactic globular clusters, multiple faint X-ray sources have been found \citep[for a review on this subject see, e.g.,][]{2006csxs.book..341V} and often multiple (candidate) quiescent neutron-star X-ray transients are found in a single globular cluster \citep[see, e.g.,][for a detailed study of quiescent transients in globular clusters]{2003ApJ...598..501H}. Therefore, it is possible that Terzan~6 might harbour several additional such sources, besides GRS\,1747--312 as well. Although the {\it XMM-Newton} position of the detected source in Terzan 6 is fully consistent with that of GRS\,1747--312, the spatial resolution of {\it XMM-Newton} is insufficient to separate potential close-by sources from each other. Therefore, despite the sub-arcsecond accuracy in the position of GRS\,1747--312 (obtained using {\it Chandra} when the source was in outburst; \citealt{intzand03}) we cannot conclusively determine if the X-ray source we detected is indeed the true quiescent counter part of GRS\,1747--312. 

Formally, we cannot even exclude the possibility that the observed emission comes from multiple sources. In particular, the different components (the hard and soft component) in the observed X-ray spectra could arise from different sources or source populations. However, this situation would give rise to multiple different scenarios. A full discussion of all these potentially scenarios is beyond the scope of the current paper since nothing conclusive can be inferred from that. Since the properties of the observed source (i.e., the two component spectrum in which both components produce roughly half the observed 0.5--10 keV X-ray luminosity, the exact shape of these two components, and the X-ray luminosity itself) are similar to several quiescent neutron-star transients, we think that it is still possible that we see emission from only one source or that it dominates the observed emission. If so, then this source is most likely the quiescent counterpart of a neutron-star X-ray transient and this could very well be GRS\,1747--312 since it is a very active X-ray transient. However, potentially it could still be another, yet unknown, system. If it is the latter case, then the next two sections (Sections \ref{subsection:ch4-accretion} and \ref{subsection:ch4-cooling}) are invalid in context of GRS\,1747--312. This source would then be much fainter than we have assumed, and we would have only an upper limit on its true quiescent emission. The neutron-star surface would then be quite cold indicating that despite the frequent outbursts of the source, the neutron-star is not significantly heated when it is accreting. This then would indicate possible enhanced neutrino emission processes that are active in the core, and thus a relatively massive neutron-star \citep[see][and discussion in Section \ref{subsection:ch4-cooling}]{colpiea01}.

If the observed X-ray emission is from an unrelated transient located in Terzan 6, we have not yet conclusively seen an outburst from this source. As shown in Figure \ref{ch4_lc}, many outbursts are observed from the direction of Terzan 6 \citep[see also][]{pavlinsky+94,intzand00,intzand03,saji16} and this trend has continued since the last one shown in the figure. Because eclipses at the expected times are seen in the data \citep[e.g.,][]{2013ATel.4915....1B,2016ATel.9072....1B}, some of these outbursts were conclusively assigned to GRS\,1747--312, although for most of the other outbursts no such observations are available so they could still, in principle, be due to another source. Moreover, \citet{saji16} reported on a {\it Suzaku} observation performed on 2009 September 16. During this observation they found a faint source ($\sim 10^{35}$ erg s$^{-1}$; faint but brighter than in quiescence, indicating transient activity from the direction of Terzan 6) but they did not find an eclipse at the expected time for GRS\,1747--312. While they discussed scenarios in which the observed emission could still have originated from GRS\,1747--312, they also provided counter arguments and therefore suggested that another source in Terzan~6 was potentially responsible for this faint emission. However, so far there is no conclusive evidence for an outburst from Terzan 6 that cannot be associated with GRS\,1747--312, although we cannot exclude that scenario either. So if the detected flux during the {\it XMM-Newton} observation indeed comes from a different source, this source has very infrequent and/or very weak outbursts (below the detectability of all-sky monitors or the PCA instrument). 

\subsection{Residual accretion}\label{subsection:ch4-accretion}

Several neutron-star transients have shown similar quiescent spectra as the one we have potentially observed for GRS\,1747--312 (e.g. \citealt{fridriksson2010,fridriksson2011,degenaar11a, degenaar15b,waterhouse16}) i.e., spectra in which the soft and hard component roughly contribute equally to the 0.5--10~keV luminosity. The exact origin of both components was for a long time not fully understood (see, e.g., the discussions in \citealt{degenaar15b}). Variability in the soft component (e.g. Cen X-4, Aql X-1; \citealt{campana97,campana04,rutledge02a,cackett10,cackett11,cackett13,bernardini13}) ruled out that the soft component was due to cooling of the neutron-star crust that is heated during outburst. Therefore, the soft component was often inferred to be due to low-level accretion onto the neutron-star surface (e.g., \citealt{campana98,campana04,rutledge02a,bernardini13}). The nature of the hard component remained elusive for many years although it had been speculated that this component (in addition to the soft component) could also be due to residual accretion onto the neutron-star surface \citep[e.g., see discussion in][]{cackett10}. 

Recently the situation has become more clear. \citet{chakrabarty14} reported on a simultaneous {\it XMM-Newton}/{\it NuSTAR} observation of the quiescent neutron-star system Cen~X--4 which showed that the hard component cuts-off at relatively low energies ($\sim$18~keV) and that this component was best described by a thermal Bremsstrahlung model. They suggested that this component could be due to emission from a hot layer just above the neutron-star surface or from a radiatively inefficient accretion flow. However, \citet{dangelo15} argued that the emission could not come from such a flow but that it had to come from this boundary layer above the neutron-star surface.

In addition, \citet{wijnands15} studied the X-ray spectra of neutron-star LMXBs that were accreting at low levels (with luminosities of $10^{34-35}$ erg s$^{-1}$), but still higher than what is observed in quiescence (when they have luminosities $<10^{34}$ erg s$^{-1}$). They noted a striking resemblance between the shape of the spectra of the quiescent systems with the systems they studied, i.e., both spectra could be described by a soft plus a hard component, and in both spectra these two components contributed roughly half to the total 0.5--10~keV luminosity. Since for systems with a luminosity of $10^{34-35}$ erg s$^{-1}$ it is clear that both components have to come from some kind of accretion process, \citet{wijnands15} suggested that this would therefore also be true for the quiescent systems that have similar two component spectra. In addition, since both components contribute roughly equally to the emission in the 0.5--10~keV energy band over a large luminosity range of $10^{32}$ to $10^{35}$ erg s$^{-1}$ (when taking into account both quiescent and weakly accreting systems), \citet{wijnands15} argued that the physical mechanisms behind both components are closely linked to each other \citep[see also][]{cackett10}, and very likely they originate very close to each other. They suggested that the energy stored in the accretion flow is liberated in such a way that half of this energy is released very close to the neutron-star (producing the hard component through Bremsstrahlung) and the other half when the matter hits the surface (producing the soft, black-body like component). This idea is consistent with the boundary layer picture presented by \citet{dangelo15}.

Since the quiescent spectrum of GRS\,1747--312 (assuming that it is the origin of all of the emission; see Section \ref{subsection:ch4-confusion} for caveat) is very similar to the spectra of the systems described in the previous two paragraphs, both in shape and in the way that both components contribute significantly to the 0.5--10~keV luminosity, we suggest that in our source as well, the emission is due to accretion onto the neutron-star (causing surface emission and emission from the boundary layer just above the surface). Previous strong evidence for residual accretion during the quiescent phase of GRS\,1747--312 was reported by \citet{intzand03}, who reported a type-I X-ray burst from this source\footnote{While the type-I burst reported by \citet{intzand03} most likely originated from a source in Terzan 6, it is not conclusively certain that this source was indeed GRS\,1747--312, and not another, unrelated, quiescent neutron-star X-ray binary in the same cluster (see Section \ref{subsection:ch4-confusion} for a discussion about potential source confusion). In this paragraph we assume that GRS\,1747--312 was indeed the source exhibiting this burst. However, if another source exhibited this type-I X-ray burst, then most of the inferences made in this paragraph about the behaviour of GRS\,1747--312 might then instead be applicable to this other source (i.e. if the quiescent emission we observed during the {\it XMM-Newton} observation also would have come fully from this other source.)} when it was in quiescence. In order for this burst to occur, the source must have accreted mass when it was already in quiescence. This burst occurred during a different quiescent interval between two outbursts than when the {\it XMM-Newton} observation was performed, suggesting that the source is frequently, or perhaps always, accreting at low levels when in quiescence. However, we note that both the quiescent type-I burst as well as the {\it XMM-Newton} observation occurred relatively close after the end of the preceding outburst ($\sim$1 month) and therefore, it remains possible that later in quiescence the source stopped fully accreting. If that is true, during a quiescent observation that would be obtained closer in time to the start of the next expected outburst, the source might show a lower quiescent luminosity and potentially also a different type of quiescent spectrum as well. True cooling emission from the neutron-star might be observable (see also Section \ref{subsection:ch4-cooling}) or we might observe a purely power-law dominated spectrum as seen in some other quiescent neutron-star transients. These options can be investigated through a quiescent monitoring campaign of the source after one of its future outbursts, preferably with a high angular resolution telescope like {\em Chandra} to avoid potential source confusion (see Section \ref{subsection:ch4-confusion}).

\subsection{Cooling of a reheated neutron-star crust}\label{subsection:ch4-cooling}

Although we consider it possible that the quiescent emission we detected is potentially contaminated (both the soft and hard component could be affected by this) by other sources in Terzan 6 (Section \ref{subsection:ch4-confusion}), it is still interesting to investigate what can be inferred if we assume that the observed soft component was due to cooling of the crust of the neutron-star in GRS\,1747--312 that was heated in outburst. Moreover, if during the outbursts the crust is significantly heated and thus has gone out of thermal equilibrium with the core, the crust might not have relaxed back to thermal equilibrium before the start of the next outburst. This would make the source very similar to Aql X-1 for which we recently investigated this possibility \cite[][]{ootes18}. Therefore, it would be interesting to compare our result for GRS\,1747--312 with those we obtained for Aql X-1. Although we note that the recurrence times of GRS\,1747--312 are similar to those of Aql~X-1, GRS\,1747--312 has shorter outbursts with lower accretion rates compared to Aql~X-1, so the influence of the accretion history might be smaller.

To investigate the evolution of the thermal state of the neutron-star in GRS\,1747--312, we have used our updated crust cooling code \textit{NSCool} \citep{pagereddy2013,page2016,ootes16} that takes into account accretion rate variability during the multiple outbursts of the source \citep[using the method outlined by][]{ootes18} to determine what we can infer about the neutron-star crust and core for our source. We modelled the full accretion rate history observed between modified Julian dates (MJDs) of 51300 (start of first outburst) and 53474 (end of the quiescent episode during which the {\it XMM-Newton} observation was performed\footnote{We did not model the history further because we are only interested in how the previous outbursts of the source would effect the state of the neutron-star during the {\it XMM-Newton} observation.}), based on the observed {\em RXTE}/PCA bulge-scan light curve (see Fig \ref{ch4_lc}). In this period, eleven accretion outbursts and quiescence episodes were observed, and the {\em XMM} observation was taken in the quiescence period after the eleventh outburst. 

In order to obtain the conversion factor for calculating the bolometric flux, we determined the maximum PCA count rate measured during each of the eleven outbursts and then obtained the mean peak count rate. We found it to be 389.9 counts\,s$^{-1}$ per 5\,PCUs. \citet{intzand00} obtained a {\em BeppoSAX} observation (using its narrow-field-instruments) during the peak of an outburst of the source and found a unabsorbed flux (0.1--200~keV) of 10.4$\times$10$^{-10}$ erg\,cm$^{-2}$\,s$^{-1}$ (see their Table 1). Assuming that this is a good approximation of the bolometric flux and that the peak flux during each outburst is roughly similar, we obtained (using the averaged PCA peak count rates) a conversion factor of 2.7$\times$10$^{-12}$ erg\,cm$^{-2}$\,s$^{-1}$ per 1 count\,s$^{-1}$ for 5\,PCUs. We used this factor to convert the observed PCA count rates into bolometric fluxes and subsequently calculate a time-dependent accretion rate which is used as input for the cooling code \citep[see][for more details about the model]{ootes18}. 

For each outburst, the start and end time of the outburst were estimated manually (see Table \ref{tab:ch4-startendtime}). To obtain the start times, we linearly extrapolated the count rates starting at the peak count rate through the data point before the peak. The start times are the times at which this extrapolation reaches zero counts s$^{-1}$. The end times were determined to be identical to the time of the first quiescent data point. At the end of outburst VI the source became Sun-constrained and hence we have no observations when the quiescent period started. Therefore, we consider the end of this outburst to be 38.33 day after the start of the outburst (38.33 days is the mean duration of the other fully sampled outbursts). 

The accretion rate was determined using

\begin{equation}
\dot{M} = \frac{F_\text{bol}4\pi D^2}{\eta c^2}
\end{equation}

\noindent
where $F_\text{bol}$ is the bolometric flux, D$=$9.5~kpc is the distance to the neutron-star, $\eta=$0.2 is the fraction of the accreted mass that is converted to X-ray luminosity and c is the speed of light. The average accretion rate for the 11 outbursts and quiescence episodes was found to be $\sim 6.6 \times$10$^{15}$ g\,s$^{-1}$, which is equal to $\sim 1 \times 10^{-10}$ $M_\odot$ per year.

To model the thermal evolution of the source, we assumed that the crustal parameters do not change between outbursts. Although this is likely not realistic for all parameters (e.g., see the studies of \citealt{deibel15,parikh17, ootes18}), the data (i.e., only one observational data point) are insufficient to take variations of these parameters into account. For consistency, we used the same mass and radius in our cooling model as used during the spectral fitting (M$=1.4M_\odot$, R$=$10~km). 

The remaining free parameters in our modelling were the core temperature ($T_0$), envelope composition ($y_\text{light}$), shallow heating depth ($\rho_\text{sh}$) and strength ($Q_\text{sh}$), and impurity parameters ($Q_{imp}$). Unfortunately, these parameters are degenerate with each other and with only one observation in quiescence we cannot firmly constrain any of these. One of the pressing issues in crust cooling research is that of the shallow heating phenomenon that is required to explain part of the sources in which crust cooling after an accretion outburst is observed (see, e.g., \citealt{wijnands17} for a review). We therefore set out to investigate whether or not it is likely that (significant) shallow heating occurs in GRS\,1747--312 as well, under the assumption that the observed quiescent emission is due to crust cooling. 

To this aim we fixed the envelope composition ($y_\text{light}=10^9\text{ g cm}^{-2}$), impurity parameter ($Q_\text{imp}=1.0$) and the range of the shallow heating depth (between 1 and $5\times10^9 \text{g\,cm}^{-3}$) to typically obtained values in other sources (e.g., \citealt{browncumming09,pagereddy2013,ootes16}). We then changed the base level of the calculated cooling curve (the observed temperature when the crust and core are in thermal equilibrium) in steps by adjusting the core temperature accordingly (note that the observed base level depends on both the core temperature and the envelope composition). For each base level we then determined the maximum shallow heating strength that is allowed by our observation. As input surface temperature we used the value obtained when we fixed the power-law photon index to 1.5 (see Table \ref{tab:ch4-specfit}). The results of our modelling are summarised in Table \,\ref{tab:ch4-nscool}. 

The results show that only if the source has a very high base level (and thus a relatively hot neutron-star core) compared to most of the other crust cooling sources (see Fig.\,2 of \citealt{wijnands17}), the presence of shallow heating can be excluded in the crust cooling scenario. If the source has a base level that is comparable to other crust cooling sources, our quiescence observation allows for a significant amount of shallow heating (1-5 MeV/nucleon) during the accretion outbursts. This value would be consistent with what is typically found in other sources and therefore GRS\,1747--312 might not be an atypical source. However, we stress that this is only valid under the aforementioned assumptions, and that we indeed see emission from a cooling hot neutron-star crust. As we have argued in Section \ref{subsection:ch4-accretion}, this is likely not true, and therefore the crust could be much colder and hence there might be no need for shallow heating at all. This final conclusion would also be true if none of the emission we observed was due to GRS\,1747--312 but to some unrelated cluster source (or sources; see Section \ref{subsection:ch4-confusion}) because the neutron-star (core and crust) in GRS\,1747--312 could then be also much colder than assumed in this section.

\begin{table}
	\centering
	\caption{Start (t$_{start}$) and end times (t$_{end}$) for each outburst used in our modelling}
	\label{tab:ch4-startendtime}
	\begin{tabular}{ccc} 
		\hline
Outburst     &t$_{start}$ &        t$_{end}$ \\ \hline
I &	51300 &	51338 \\
II &	51444 &	51478 \\
III &	51584 &	51623 \\
IV &	51755 &	51803 \\
V &	52059 &	52090 \\
VI &	52195 &	52233 \\
VII &	52328 &	52367 \\
VII &	52465 &	52500 \\
IX &	52695 &	52747 \\
X & 	53074 &	53116 \\
XI &	53222 &	53260 \\ \hline
\end{tabular}
\end{table}

\begin{table}
	\centering
	\caption{The results obtained using our \textit{NSCool} model}
	\label{tab:ch4-nscool}
	\begin{tabular}{lcc} 
		\hline
$T_0$ 	&	Base level 	&	Maximum Q$_{sh}$ \\
($10^8$ K) & (eV) & (MeV nucleon$^{-1}$)\\
\hline
1.58	&	$\sim$120			&	0\\
1.43	&	$\sim$115			&	1.05\\
1.28	&	$\sim$110			&	2.15	\\
1.13	&	$\sim$105			&	3.10\\
1.00	&	$\sim$100			&	4.00\\
0.890	&	$\sim$95			&	4.75\\
0.785	&	$\sim$90			&	5.45\\ \hline
\multicolumn{3}{l}{Note: All parameters (including shallow heating strength) }\\
\multicolumn{3}{l}{are kept constant between the different outbursts. See text}\\
\multicolumn{3}{l}{for more details and the exact values of these parameters.}\\
\end{tabular}
\end{table}

\section*{Acknowledgements}

SV would like to acknowledge support from NOVA (Nederlandse Onderzoekschool Voor Astronomie). RW, AP and LO acknowledge support from a NWO Top Grant, module 1, awarded to RW. ND is supported by an NWO VIDI Grant. DP is supported by Grant No. 240512 from Conacyt CB-2014-1.




\bibliographystyle{mnras}
\bibliography{biball.bib} 






\bsp	
\label{lastpage}
\end{document}